\documentclass[aps,prb,superscriptaddress,amsmath,amssymb,floatfix,twocolumn,showpacs]{revtex4-1}
\usepackage{graphicx,graphics,color}
\usepackage{dcolumn}
\usepackage{bm}

\newcommand{\up}{\uparrow}
\newcommand{\dn}{\downarrow}
\newcommand{\beq}{\begin{equation}}
\newcommand{\eeq}{\end{equation}}
\newcommand{\beqn}{\begin{eqnarray}}
\newcommand{\eeqn}{\end{eqnarray}}

\begin{document}
\preprint{submitted to Phys. Rev. B}

\title{Explicit inclusion of electronic correlation effects in molecular dynamics}
\author{Jean-Pierre Julien}
\affiliation{CNRS/Universit\'{e} de Grenoble Alpes-Institut N\'{e}el, France}
\author{Joel D. Kress}
\affiliation{Theoretical Division, Los Alamos National Laboratory,
Los Alamos, New Mexico 87545}
\author{Jian-Xin Zhu}
\affiliation{Theoretical Division, Los Alamos National Laboratory,
Los Alamos, New Mexico 87545}
\affiliation{Center for Integrated Nanotechnologies, Los Alamos National Laboratory,
Los Alamos, New Mexico 87545}

\date{\today}
\begin{abstract}
We design a quantum molecular dynamics method for strongly correlated electron metals. The strong electronic correlation effects are treated within a real-space version of the Gutzwiller variational approximation (GA), which is suitable for the inhomogeneity inherent in the process of quantum molecular dynamics (MD) simulation.  We also propose an efficient algorithm based on the second-moment approximation to the electronic density of states for the search of the optimal variation parameters,  from which the effective renormalized interatomic MD potentials are fully determined. By considering a minimal one-correlated-orbital Anderson many-particle model based on tight-binding hopping integrals, this fast GA-MD method is benchmarked with that using exact diagonalization to solve the GA variational parameters. In the infinite damping limit,  the efficiency and accuracy are illustrated. This novel method will open up an unprecedented  opportunity enabling large-scale quantum MD simulations of strongly correlated electronic materials.
\end{abstract}

\maketitle
\section{Introduction}
Electronic correlation effects in materials such as transition metal oxides, give rise to emergent phenomena including Mott insulator, magnetism, heavy fermion, and unconventional superconductivity. These phenomena defy the description of  the density functional theory (DFT) within local density approximation (LDA), which has been successful in describing electronic and structural properties of good metals and several semiconductors.  Other discrepancies show up for materials like elemental actinide solids.  For instance, the equilibrium volume of $\delta$-plutonium is experimentally 25\% larger than the one given by the DFT-LDA approach, the greatest deviation known between experiment and theoretical value for this theory. The inadequacy of the DFT-LDA method for strongly correlated electron materials can be partly cured  by including a direct treatment of quantum fluctuation effects by such quantum many-body approaches like the dynamical mean-field theory
(DMFT).~\cite{AGeorges96,GKotliar06} Together with its success in describing key physical observables in many strongly correlated electron materials,  however, the LDA+DMFT is computationally expensive and in practice limited to solid state systems with high crystalline symmetry, making it time consuming to describe the structural relaxation problems.   The combination of LDA with the Gutzwiller variational method~\cite{Gutzwiller1963,Gutzwiller1965} has proved successful in providing an alternative but fast approach to the strongly correlated electron metals.~\cite{Julien2006}  When combined with the MD, the computational efficiency of the GA method makes it ideal for the studies of such problems as material structure stability in strongly correlated electron metals.

The strategy  seems to be straightforward in principle. However,
practical MD simulations for low symmetry structures (such as
defects, surfaces, clusters, and liquids) containing thousands of
atoms (and accompanying electrons) present challenges. On the one
hand, for an explicit treatment of strong electronic correlation
effects, the ab initio method requires a definition of local
correlated orbitals.  On the other hand, interatomic forces
derived from the correlated wavefunctions need to be calculated
rapidly and repeatedly during the time evolution of the MD
trajectory. The aim of this work is to present a generic framework
of the GA-MD method, together with a path forward for improving
the computational speed of the GA optimization procedure. We
propose the construction of such parameterizations as presented in
the tight-binding electronic structure method, and the use of  the
semi-empirical second moment approximation of electronic density
of states for the calculation of local kinetic energy. The latter
will significantly speed up the minimization procedure in the
Gutzwiller variational method for the electronic structure,
opening up the possibility of MD simulations to strongly
correlated electronic materials.

The outline of the paper is as follows. In Sec.~\ref{sec:GA}, we
give a detailed description of the density matrix formulation of
the Gutzwiller approximation. It has the advantage of being
applicable to crystals, as well as topologically and/or chemically
disordered systems.   In Sec.~\ref{sec:variational}, we derive an
approximate but analytical solution to the optimization equations
in the GA, where a high quality fitted solution on the whole range
of physical interest is provided, and propose the second moment
approach to the electronic density of states for the calculation
of kinetic energy parameters.
 In Sec.~\ref{sec:model}, this efficient GA-MD method is demonstrated in a minimal Anderson model for heavy fermion systems based on tight-binding hopping integrals. A concluding summary is given in Sec.~\ref{sec:summary}.

\section{Density matrix formulation of Gutzwiller method}
\label{sec:GA}
\subsection{Renormalization of hopping integrals}
First, we review the Gutzwiller method briefly for which we
closely follow Ref. \onlinecite{Julien2006} but here we specifically include the topological disorder, i.e.,
during a typical MD process, no symmetry remains and all atoms are inequivalent.
Among numerous theoretical approaches, the Gutzwiller method provides a transparent physical
interpretation in terms of the atomic configurations of a given
site. Originally, it was applied to the one-band Hubbard model
Hamiltonian:~\cite{Hubbard1963}
\begin{equation}
H = H_{kin} + H_{int}\;,
\label{hamiltonian}
\end{equation}
with
\begin{equation}
H_{kin}=\sum_{i \neq j, \sigma} t_{ij}c_{i\sigma}^\dagger
c_{j\sigma}\;,
\end{equation}
and
\begin{equation}
H_{int}= U\sum_i n_{i\uparrow}n_{i\downarrow}\;.
\end{equation}
The Hamiltonian contains a kinetic part $H_{kin}$ with a hopping
integral $t_{ij}$ from site $j$ to $i$, and an interaction part
with a local Coulomb repulsion $U$ for electrons on the same site.
$c_{i\sigma}^\dagger$ ($c_{j\sigma}$) is the creation
(annihilation) operator of an electron at site $i$ with up or down
spin $\sigma$. $n_{i\sigma}=c_{i\sigma}^\dagger c_{i\sigma}$
measures the number (0 or 1) of electron at site $i$ with spin
$\sigma$. The Hamiltonian, Eq.~(\ref{hamiltonian}), contains the key
ingredients for correlated up and down spin electrons on a
lattice: the competition between delocalization of electrons by
hopping and their localization by the interaction. It is one of the
most widely used models to study the electronic correlations in solids.

In the absence of the interaction $U$, the ground state is
characterized by the Slater determinant comprising the
Hartree-like wave functions (HWF) of the uncorrelated electrons,
$|\psi_0\rangle$. When $U$ is switched on, the weight of the doubly
occupied sites will be reduced because of  the cost of an additional
energy $U$ per site. Accordingly, the trial Gutzwiller wave
function (GWF) $|\psi_G\rangle$ is built from the  HWF
$|\psi_0\rangle$,
\begin{equation}
|\psi_G\rangle = g^D|\psi_0\rangle.
\end{equation}
The role of $g^D$ is to reduce the weight of the configurations
with doubly occupied sites, where $D=\sum_i
n_{i\uparrow}n_{i\downarrow}$ measures the number of double
occupations and $g\; (<1)$ is a variational parameter. This
method corrects the mean field (Hartree) approach, for which up and
down spin electrons are independent, and,  overestimates
configurations with doubly occupied sites. Using the Rayleigh-Ritz
principle, this parameter is determined by minimization of the
energy in the Gutzwiller state $|\psi_G\rangle$, giving an upper
bound to the true unknown ground state energy of $H$. To
enable a practical calculation, it is necessary to use
the Gutzwiller  approximation, which assumes that all
configurations in the HWF have the same weight.

Nozieres~\cite{Nozieres1986} proposed an alternative way which
shows that the Gutzwiller approach is equivalent to the
renormalization of the density matrix in the GWF. It can be
formalized as
\begin{equation}
\rho_G = T^\dagger \rho_0 T\;.
\label{renormalizeddensitymatrix}
\end{equation}
The density matrices $\rho_G=|\psi_G\rangle \langle \psi_G |$ and
$\rho_0 = |\psi_0 \rangle \langle \psi_0 |$ are projectors on the
GWF and HWF, respectively. $T$ is an operator which is diagonal in
the configuration basis; $T=\Pi_i T_i$ where $T_i$ is a diagonal
operator acting on site $i$,
\begin{equation}
T_i|L_i,L' \rangle=\sqrt{\frac{p(L_i)}{p_0(L_i)}}|L_i,L' \rangle.
\label{tioperator}
\end{equation}
Here, $L_i$ is an atomic configuration of the site $i$, with
probability $p(L_i)$ in the GWF and $p_0(L_i)$ in the HWF
respectively, whereas $L'$ is a configuration of the remaining
sites of the lattice. Note that this prescription does not change
the phase of the wave function as the eigenvalues of the operators
$T_i$ are real. The correlations are local, and the configuration
probabilities for different sites are independent.

The expectation value of the Hamiltonian is given by,
\begin{equation}
\langle H \rangle_G = \text{Tr}(\rho_G H)\;.
\label{hamiltonian_gutzwiller}
\end{equation}
The mean value of the on-site operators is exactly calculated with
the double occupancy probability, $d_i=\langle
n_{i\uparrow}n_{i\downarrow} \rangle_G$. Therefore, $d_i$ is the new
variational parameters replacing $g$. Using Eqs.~(\ref{renormalizeddensitymatrix})-(\ref{tioperator}), the two-site
operator contribution of the kinetic energy can be written as,
\begin{widetext}
\begin{equation}
\langle c_{i\sigma}^\dagger c_{j\sigma}\rangle_G = \text{Tr}(\rho_G
c_{i\sigma}^\dagger c_{j\sigma})=\langle c_{i\sigma}^\dagger
c_{j\sigma}
\rangle_0\sum_{L_{-\rho}}\sqrt{\frac{p(L'_\sigma,L_{-\sigma})}{p_0(L'_\sigma)}}
\sqrt{\frac{p(L_\sigma,L_{-\sigma})}{p_0(L_\sigma)}}\;,
\end{equation}
\end{widetext}
where $L'_\sigma$ and $L_\sigma$ are the only two configurations
of spin $\sigma$ at sites $i$ and $j$ that give a non-zero matrix
element for the operator in the brackets. The summation is
performed over the configurations of opposite spin $L_{-\sigma}$.
The probabilities $p_0$ in the HWF depend only on the number of
electrons, whereas the $p$ in the GWF also depends on $d_i$.

After some elementary algebra, one can show that the Gutzwiller
mean value can be factored into,
\begin{equation}
\langle c_{i\sigma}^\dagger
c_{j\sigma}\rangle_G=\sqrt{q_{i\sigma}} \langle
c_{i\sigma}^\dagger c_{j\sigma}\rangle_0 \sqrt{q_{j\sigma}},
\label{normalizedenergy}
\end{equation}
where these renormalization factors $q_{i\sigma}$ are local and
can be expressed as
\begin{equation}
\sqrt{q_{i\sigma}}=\frac{ \sqrt{1-n_{i\sigma}-n_{i-\sigma}+d_i} \sqrt{n_{i\sigma}-d_i}
+ \sqrt{d_i} \sqrt{n_{i-\sigma}-d_i}}{\sqrt{n_{i\sigma}(1-n_{i\sigma})}} \;.
\label{qi}
\end{equation}
In Eq.~(\ref{normalizedenergy}), $ \langle
c_{i\sigma}^\dagger c_{j\sigma}\rangle_0$ is shorthand for the expectation value of  $c_{i\sigma}^\dagger c_{j\sigma} $ over the HWF  $|\psi_0\rangle$, that is, $\langle \psi_0| c_{i\sigma}^\dagger c_{j\sigma} |\psi_0\rangle$,
and similarly  for the average  over the Gutzwiller state $\Psi_{G}\rangle$.
We have also used $n_{i\sigma}$ as shorthand  for $\langle
n_{i\sigma} \rangle$, that is, the average number of electrons on the
considered ``orbital-spin'' in the HWF.
In the simple case when the state is homogeneous and paramagnetic, all quantities becoming site-
 and spin-independent.

In Eq.~(\ref{normalizedenergy}), the term contributing to the kinetic energy, $ \langle c_{i\sigma}^\dagger
c_{j\sigma}\rangle_0 $, is renormalized by a
factor of $q$, which is less than one in the correlated state,
and equal to one in the HWF. This factor can be interpreted as a
direct measure of the correlation effect. Indeed Vollhardt~\cite{Vollhardt1984} has shown that $1/q=m^*/m$ where $m^*$ is the
effective mass and $m$ is the bare mass of the electron. Thus a
$q$ close to $1$ corresponds to a weakly correlated electron system
and a smaller $q$ value reflects enhancement of the correlation
effect. Equation~(\ref{hamiltonian_gutzwiller}) leads to the variational
energy per site, and for the homogeneous and paramagnetic state, is given by
\begin{equation}
E(d)=\langle H \rangle_G=2q \varepsilon^0_{kin} + Ud \;,
\end{equation}
which can be minimized numerically with respect to the variational parameter
$d$. In the above expression, the factor 2 accounts for the two-fold spin degeneracy and   $\varepsilon^0_{kin}$ is the kinetic energy per site and per spin identical at all sites and spins for an homogeneous HWF,
\begin{equation}
\varepsilon^0_{kin}=\sum_{j}  \langle c_{i\sigma}^\dagger c_{j\sigma}\rangle_0 t_{ij}
\end{equation}
In the case of half filling ($n=1/2$), minimization is analytical, and provides
the optimal choice for double occupancy $d$
\begin{equation}
d =\frac{1}{4}\biggl{(}1-\frac{U}{16 \varepsilon^0_{kin}}\biggr{)}\;,
\end{equation}
and
\begin{equation}
q = 1-\frac{U^2}{(16 \varepsilon^0_{kin})^2}\;.
\end{equation}
If the Coulomb repulsion $U$ exceeds a critical
value $U_c = 16 \varepsilon^0_{kin}$, $q=0$,
leading to an infinite quasiparticle mass with a Mott-Hubbard
Metal-Insulator transition. This is also known as ``the
Brinkmann-Rice transition'',~\cite{Brinkmann1970} as these authors
first applied the Gutzwiller approximation to the Metal-Insulator
transition.

Away from half-filling, one has to minimize the variational energy
of  Eq.~(\ref{hamiltonian_gutzwiller}) numerically. Moreover if the system is inhomogeneous,
which is the case for a MD simulation,  all quantities ($d_i$, $q_i$)
may vary locally from one site to the other. Consequently, the general
variational energy, a function of double occupancy probabilities $d_i$ on all sites,
is
\begin{equation}
E_{var}=  \sum_{ij\sigma} \sqrt{q_{i\sigma}} t_{ij}\sqrt{q_{j\sigma}} \langle c_{i\sigma}^\dagger c_{j\sigma}\rangle_{0} + \sum_{i} U_i d_i
\end{equation}

Minimization must then be performed numerically for each site, i.e.,
derivation with respect to  $d_i$ , leading to the local equation:
\begin{equation}
\label{miniz-equa}
\frac{\partial \sqrt{q_{i\sigma}}}{\partial d_{i}}= \frac{U_i}{4|e_{i\sigma}|}\;.
\end{equation}
Here $e_{i\sigma}$  is the local partial ``effective'' kinetic energy, i.e., the
contribution of orbital-spin ``$i\sigma$'' to kinetic energy, but calculated with  an ``effective'' hopping,
 renormalized by $q$,
\begin{equation}
e_{i\sigma} =   \sum_{j} \langle c_{i\sigma}^\dagger c_{j\sigma}\rangle_0 t_{ij} \sqrt{q_{j\sigma}}
\end{equation}

\subsection{Inequivalent sites: renormalization of levels}
When sites are inequivalent, or if orbitals belong to different
symmetries as in a multiorbital basis, it is necessary
to add to the Hamiltonian an on-site
energy term
\begin{equation}
H_{on-site}=\sum_{i\sigma}\epsilon^0_{i\sigma}n_{i\sigma}
\end{equation}
The Hubbard Hamiltonian is written
as
\begin{equation} \label{hub+onsite}
 H=\sum_{i\neq j,\sigma} t_{ij}
c_{i\sigma}^\dagger
c_{j\sigma}+\sum_{i\sigma}\epsilon^0_{i\sigma}n_{i\sigma}+U\sum_i
n_{i\up} n_{i\dn}
\end{equation}
In this case, the starting HWF directly obtained from
the non-interacting part of the Hamiltonian, is not automatically
the optimal choice, i.e., having the lowest
energy.
For example, if we look for the ground state of Eq.~(\ref{hub+onsite}) in the Hartree-Fock (HF) self-consistent field
formalism, it is necessary to vary the orbital occupations.
Practically, it can be achieved by replacing Eq.~(\ref{hub+onsite}), by
an effective Hamiltonian $H_{eff}$ of independent particles with
renormalized on-site energies $\epsilon_{i\sigma}$:
\begin{equation} \label{Heff} H_{eff}=\sum_{i\neq j,\sigma} t_{ij}
c_{i\sigma}^\dagger
c_{j\sigma}+\sum_{i\sigma}\epsilon_{i\sigma}n_{i\sigma}+C\;.
\end{equation}

The HWF we are looking for is an \textit{approximate} ground
state of the \textit{true} many-body Hamiltonian
(\ref{hub+onsite}) and is the \textit{exact} ground state of
\textit{effective} Hamiltonian (\ref{Heff}). The additive constant
$C$ accounts for double counting energy reference, so that the
ground state energies are the same for both Hamiltonians:
\begin{equation} \label{heff=h}
\langle H_{eff} \rangle=\langle H \rangle\;.
\end{equation}
The optimal choice of parameters $\epsilon_{i\sigma}$ can be obtained by
minimizing the ground state energy of $H_{eff}$ with respect to
$\epsilon_{i\sigma}$ . With the help of Hellmann-Feyman theorem, one
can obtain the derivative of the kinetic energy
\begin{equation} \label{derivHkin}
\frac{\partial\langle H_{kin}\rangle}{\partial\epsilon_{i\sigma}}=
-\sum_{j\neq i,\sigma}\epsilon_{j\sigma}\frac{\partial\langle
n_{j\sigma}\rangle}{\partial\epsilon_{i\sigma}}  \;.
\end{equation}
On the other hand, differentiation of Eq.~(\ref{heff=h})  in association with Eq.~(\ref{derivHkin}) and
the mean field approximation  $\langle n_{i\up}
n_{i\dn}\rangle\approx\langle n_{i\up}\rangle\langle
n_{i\dn}\rangle $ recovers the well-known formula for
the on-site energies
\begin{equation}
\epsilon_{i\sigma}=\epsilon_{i\sigma}^{0}+ U \langle
n_{i-\sigma}\rangle \;,
\label{renoHF1}
\end{equation}
where the constant $C$ is simply
$-U\sum_i\langle
n_{i\up}\rangle\langle n_{i\dn}\rangle $.

In the Gutzwiller approach, the same argument about the variation
of orbital occupation, i.e., flexibility on the HWF
$|\psi_0\rangle$, is true. It is necessary to find a way to vary
this Slater determinant, from which the GWF $|\Psi_G\rangle$ is
generated, so that the Gutzwiller ground-state energy is a minimum.
One needs to find an equivalent of Eq.~(\ref{renoHF1}) in
the Gutzwiller context. The average value of Eq.~(\ref{hub+onsite}) on
a GWF is given by:
\begin{eqnarray}
\label{average-hub+onsite}
\langle
\Psi_G|H|\Psi_G\rangle&=&\sum_{ij\sigma}t_{ij}\sqrt{q_{i\sigma}}\langle
c_{i\sigma}^\dagger c_{j\sigma}\rangle_0\sqrt{q_{j\sigma}}+U\sum_i
d_i\nonumber \\
&&+\sum_{i\alpha\sigma}\epsilon^0_{i\sigma}\langle
n_{i\sigma}\rangle_0 \;.
\end{eqnarray}
Following the previous HWF self-consistent field
approach, one can  find an effective Hamiltonian $H_{eff}$ of
independent particles having $|\psi_0\rangle$ as an \textit{exact}
ground state. This state $|\psi_0\rangle$ generates the GWF
$|\Psi_G\rangle$ which is an \textit{approximate} ground state of
the true  Hamiltonian Eq.~(\ref{hub+onsite}). In analogy
with Eq.~(\ref{heff=h}),
\begin{equation}
\label{heff=h2}
\langle \psi_0|H_{eff}|\psi_0\rangle=\langle
\Psi_G|H|\Psi_G\rangle\;,
\end{equation}
leads to the expression:
\begin{equation}
\label{Heff2}
H_{eff}=\sum_{i\neq j,\sigma} \tilde{t}_{ij}
c_{i\sigma}^\dagger
c_{j\sigma}+\sum_{i\sigma}\epsilon_{i\sigma}n_{i\sigma}+C^{\prime} \;,
\end{equation}
 with effective but \textit{fixed} renormalized hopping integrals
$\tilde{t}_{ij}=\sqrt{q_{i\sigma}} t_{ij} \sqrt{q_{j\sigma}}$ and
effective on-site energies $\epsilon_{i\sigma}$,
still to be determined. The Hellmann-Feynman theorem applied to
$H_{eff}$ provides again an expression similar to
Eq.~(\ref{derivHkin}), but with effective hopping integrals. Taking into
account the dependence of the ${q_{i\sigma}}$'s through
${n_{i\sigma}}$ (Eq.~(\ref{qi})) and differentiating
Eqs.~(\ref{average-hub+onsite}) and
(\ref{heff=h2}) with respect to the parameters
$\epsilon_{i\sigma}$,  one obtains the
equivalent expression to Eq.~(\ref{renoHF1}) in the Gutzwiller context:
\begin{equation} \label{renoG}
\epsilon_{i\sigma}=\epsilon^0_{i\sigma}+2e_{i\sigma}\frac{\partial
\ln(\sqrt{q_{i\sigma}})}{\partial n_{i\sigma}}  \;.
\end{equation}
Here $e_{i\sigma}$ is the partial kinetic energy of orbital-spin
$i \sigma$
\begin{equation}
\label{partial-kin}
e_{i\sigma}=\sum_{j\sigma}\tilde{t}_{ij}
\langle c_{i\sigma}^\dagger
c_{j\sigma}\rangle_0=\int_{-\infty}^{\rm E_F}
E\tilde{N}_{i\sigma}(E)dE-\epsilon_{i\sigma}\langle
n_{i\sigma}\rangle_0  \;,
\end{equation}
with $\tilde{N}_{i\sigma}$ the $i \sigma$-projected density of
states (DOS) for a system described $H_{eff}$. Equation~(\ref{heff=h2})
leads to
\begin{equation}
C^{\prime}=U\sum_i d_i-\sum_{i\sigma} 2e_{i\sigma}\frac{\partial
\ln(\sqrt{q_{i\sigma}})}{\partial n_{i\sigma}}\langle n_{i\sigma}\rangle\;.
\end{equation}

Except for a few very special conditions in one-band Hubbard model,
the renormalization of correlated-orbital levels is not only important in the optimization
of the total energy but  also in giving a
correct description of single-particle quasiparticle properties.~\cite{JXZhu2012}
To solve the full problem of finding an approximate ground state
to Eq.~(\ref{hub+onsite}), one is faced with a
self-consistency loop: First get the occupations $\langle n_{i\sigma}\rangle_0$ from a
HWF, and a set of `bare' $\epsilon^{0}_{i\sigma}$ levels; then
obtain  a set of configuration parameters, the probabilities of
double occupation, $d_i $ by minimizing Eq.~(\ref{average-hub+onsite})
with respect to these probabilities, followed by the on-site level renormalization
 according to Eq.~(\ref{renoG}). The loop repeated until
a convergence is achieved.


\section{Approximated solutions of minimization equations}
\label{sec:variational}
\begin{figure}[h]
\includegraphics[scale=0.4]{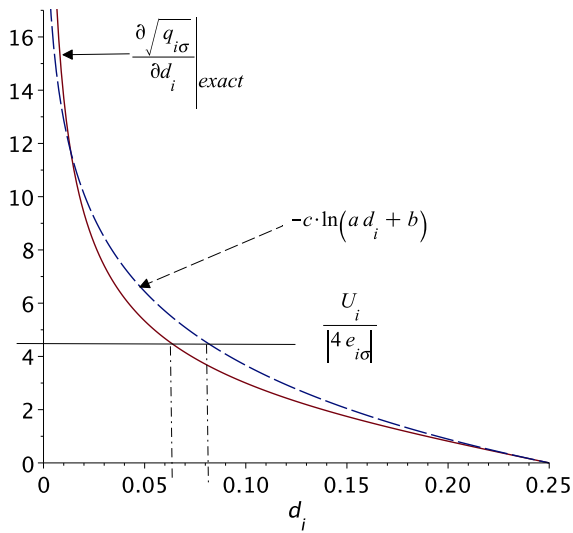}
\caption{(Color online) The exact (full line) and approximate (dashed line) $\partial \sqrt{q_{i\sigma}}/\partial d_i$  as a function of $d_i$.}
\label{FIG:root}
\end{figure}

Due to the complicated expression of  Eq.~(\ref{qi}),  it is non-trivial to solve Eq.~(\ref{miniz-equa}). Graphically, the solution corresponds to
the intersection of the function  $\frac{\partial \sqrt{q_{i\sigma}}}{\partial d_i}$  with a
horizontal line $U/4 |e_{i\sigma}|$  (see Fig.~\ref{FIG:root}).

This situation may seem
insurmountable for application of  the Gutzwiller method to MD, as
analytical expressions are necessary to be able to derive forces.
Fortunately, the function $\frac{\partial \sqrt{q_{i\sigma}}}{\partial d_{i}}$  can be fitted with
reasonable  accuracy (see Fig.~\ref{FIG:root})  by a logarithm function,  giving
an analytical approximate solution $d_i$ of Eq.~(\ref{miniz-equa}).  This
choice was suggested by the shape of the true  derivative  of  $\sqrt{q_{i\sigma}}$ ,
keeping in mind the following physical constraints:
The uncorrelated case ($U=0$) has to give the solution $d_i= n_{i\sigma}^{2}$, for a
given occupancy $n_{i\sigma}$,  and the  probability of double occupancy $d_i$ is
restricted in the range  $\text{max}(0, 2n_{i\sigma}-1) < d_i < n_{i\sigma}$
(otherwise there would be negative arguments in the square root  of $q$),
providing a rescaling of the logarithm argument. Finally, we adapt the coefficient in
front of the logarithm, in such a way that the fitted function
has the same slope as the true one in the uncorrelated
limit $n_{i\sigma}^2$. The final result reads:
\begin{equation}
\frac{\partial \sqrt{q_{i\sigma}}}{\partial d_i}\simeq -c\ln(ad_i+b)
\label{approx-deriv-sqrtq}
\end{equation}
The physical constraints above fix uniquely all three coefficients
\begin{eqnarray}
a &=& \frac{1}{n_{i\sigma}^2-\text{max}(0,2n_{i\sigma}-1)}\;,\\
b &=& -a\; \text{max}(0,2n_{i\sigma}-1) \;,\\
c &=& \frac{n_{i\sigma}^{2}-\text{max}(0,2n_{i\sigma}-1)}{4n_{i\sigma}^{3}(1-n_{i\sigma})^{3}}\;.
\end{eqnarray}

The approximate value of double occupancy, the solution of minimization equation, within this approximation, is
$d_i=(n_{i\sigma}^2-d_m) \exp(-U_i/|4 e_{i\sigma}|)-d_m$, where
$d_m =\text{max}(0,2n-1)$. The small remaining difference between this approximate and the true value
 can be corrected by a second order expansion around the approximate value $d_i$
leading to an accurate  analytical expression
\beq
d_i^{2nd}=d_i-\frac{f'+\sqrt{f'^2-2f''[f + c\ln(ad_i+b)]}}{f''}\;.
\eeq
Here $f$, $f'$, and $f''$ stand for the true $\frac{\partial \sqrt{q_{i\sigma}}}{\partial d_i}$, and its first and second order derivatives, respectively,  calculated at the approximated value $d_i$.

The relative error on this second order corrected value with respect to the exact solution is less than 1\% over the whole
range of values. This second order corrected local double occupancy is the one now used  in the calculation of
renormalization factor $\sqrt{q_{i\sigma}}$, Eq.~(\ref{qi}).
To check the validity of this approximation for the
derivative, we also plot in Fig.~\ref{FIG:root} the comparison between true and
approximate $\sqrt{q_{i\sigma}}$. Again we see the good accuracy, the small
discrepancy being in the range of very small double occupancy,
i.e., corresponding to high value of Coulomb repulsion $U$, which
are far from the values for realistic materials.

%

The other input for Eq.~(\ref{partial-kin})  necessary to perform tractable MD simulations is the partial energy. A common approximation
is the well-known approach of second moments~\cite{JFriedel:1964} with the assumption of a rectangular electronic density of states of bandwidth
$W_{i\sigma}$ and height $1/W_{i\sigma}$ for each spin.
The resulting second moment of the rectangular band is $\mu_{2,i}=W^2_{i\sigma}/12$.
For a given MD snapshot, the second moment for a given atomic site ``$i$'' can be constructed as a sum over atoms ``$j$''
neighboring ``$i$'' as $\mu_{2,i} = \sum_{j} t_{ij}^2$ (see Appendix for more details).
Using the simple tight-binding theory,~\cite{WAHarrison:1980} the hopping integrals $t_{ij}$ scale as
a power law of the interatomic distance $r_{ij} =\vert \mathbf{r}_i -\mathbf{r}_j\vert$.
With the number of  electrons on each atom (assuming charge neutrality
  for all sites),  the partial kinetic energy needed as input in Eq.~(\ref{partial-kin})
  and is given by
$e_{i\sigma}= W_{i\sigma} n_{i\sigma}(n_{i\sigma}-1)/2$,  similar to the result by Ackland.~\cite{Ackland2003}
To compute  $\mu_2$  and to account for the effect
of Coulomb correlations, we use the hopping integrals,
renormalized by $q$-factors, $\sqrt{q_i}   t_{ij} \sqrt{q_j}$, rather than the bare
ones, $t_{ij}$: After minimization in the Gutzwiller method,  the true
interacting Hamiltonian $H$, is replaced by an effective Hamiltonian of
non-interacting quasiparticles, with renormalized hopping integrals
(and potentially renormalized on-sites too, but it is useless in
our case, as we assumed charge neutrality, so there are no average
charge transfer between sites). To conclude, we see that inclusion
of electronic correlations in MD simulations within the Gutzwiller
method, just requires one more intermediate step, compared to usual
one, for having the renormalization factors that reduce a bit the
hopping integrals. Once they have been computed  for a set of
actual positions of atoms,  the rest of the process is similar to
the usual way developed in other semi-empirical approaches.~\cite{MFinnis:1984,MSDaw:1983}

\section{Model and results}
\label{sec:model}
\subsection{Model}
To illustrate the possibility of the method, we consider  a minimal two-orbital model that mimics, e.g., heavy fermions or actinides systems, with one non-correlated band, called for convenience ``d'', whereas the other, called ``f'',  bears a strong local Coulomb repulsion $U$. For simplicity, each of these two orbitals has a spin-$\frac{1}{2}$ degree of freedom per site. This model is described by the following Hamiltonian (close in spirit to the Anderson lattice model) with the usual notation:
\begin{eqnarray}
H&=&\sum_{i,j,\sigma} t_{i \textit{d}, j \textit{d}} c_{i\textit{d}\sigma}^\dagger c_{j\textit{d}\sigma}
+t_{i \textit{d}, j \textit{f}} c_{i\textit{d}\sigma}^\dagger c_{j\textit{f}\sigma}
+t_{i \textit{f}, j \textit{d}} c_{i\textit{f}\sigma}^\dagger c_{j\textit{d}\sigma} \nonumber \\
&&+ \sum_{i\sigma}(\epsilon^0_{i \textit{d}\sigma}n^\textit{d}_{i \sigma}+\epsilon^0_{i \textit{f} \sigma}n^\textit{f}_{i\sigma}) \nonumber \\
&&+\sum_iU_i n^\textit{f}_{i\up}n^\textit{f}_{i\dn} \;.
\label{H-model}
\end{eqnarray}
Here the $d$-orbitals are coupled among themselves and with $f$-orbitals, whereas the $f$-orbitals are only coupled to their neighboring $d$-orbitals.
The power laws~\cite{WAHarrison:1984} in distance from atom located at position $r_i$ to atom at $r_j$ for hopping integrals are -5 and -6 for \textit{dd}-coupling and \textit{df}-coupling, respectively:
 \begin{equation}
 t_{i \textit{d}, j \textit{d}} =  \frac{t_{dd,0}}{|r_i-r_j|^5}\;,
 \label{powerlaw}
\end{equation}
and
\begin{equation}
 t_{i \textit{d}, j \textit{f}} = \frac{t_{df,0}}{|r_i-r_j|^6}\;,
\label{powerlaw}
\end{equation}
where $t_{dd,0}$ and $t_{df,0}$ are constants.
  After a Gutzwiller  mean-field treatment of Hamiltonian Eq.~(\ref{H-model}), we obtain
\begin{eqnarray}
H_{eff}&=&\sum_{i,j,\sigma} t_{i \textit{d}, j \textit{d}} c_{i\textit{d}\sigma}^\dagger c_{j\textit{d}\sigma}
+\sum_{i,j,\sigma} [t_{i \textit{d}, j \textit{f}}\sqrt{q_j} c_{i\textit{d}\sigma}^\dagger c_{j\textit{f}\sigma} +\text{H.c.}] \nonumber \\
&&+ \sum_{i\sigma}(\epsilon^0_{i \textit{d}\sigma} n^\textit{d}_{i \sigma}+\epsilon_{i \textit{f} \sigma}n^\textit{f}_{i\sigma}) +\sum_i U_i d_i + C^{\prime}\;,
\label{Heff-model}
\end{eqnarray}
whose parameters are obtained from the minimization procedure analogous to deriving Eq.~(\ref{Heff2}) from Eq.~(\ref{heff=h2}), for a given set of atomic positions. When the converged Gutzwiller ground state has been obtained we calculate the forces on each atom. These attractive forces have a quantum origin, due to the hybridization through hopping integrals. To mimic the short range repulsion that accounts for  the Pauli principle when atoms get too close,  we add a phenomenological repulsive potential
\begin{equation}
E_{\text{rep}}=\frac{1}{2} \sum_{ij} \frac{\Lambda_0}{\vert \mathbf{r}_i -\mathbf{r}_j\vert^{12}}
\label{Erep}
\end{equation}
with $\Lambda_0$ a constant.

\subsection{Calculation of forces}
For a given set of atomic positions, the overall total energy of
the system is the sum of the electronic approximate Gutzwiller
ground state energy $E_G$ plus the short range repulsion
potential, \beq E_{tot}=E_G+E_{\text{rep}}\;.
\label{eq:totalenergy} \eeq The $x$- component of the force acting
on atom $i$, $F_{x,i}$ is the derivative of $E_{tot}$ with respect
to position component $x_i$ of this atom (same relations hold for
$y$- and $z$-components):
\beq
\label{eq:fxx}
 F_{x,i}=-\frac{\partial E_{tot}}{\partial x_i}= -\frac{\partial
E_G}{\partial x_i}-\frac{\partial E_{\text{rep}}}{\partial x_i}\;.
\eeq
 From  the Hellmann-Feynman theorem, the first term, due to hybridization, can  be split
 into elementary contributions:
\beq
-\frac{\partial E_{G}} {\partial x_i}  =  \sum_{j \neq i} \sum_{\alpha \beta} f^{(x)}_{i \alpha j \beta}
\eeq
where the contribution of orbitals $\alpha$ of site $i$ and $\beta$ of site $j$ ($\alpha$ or $\beta$ are either $d$- or $f$-orbitals) is related to the
derivative of the hopping integral $t_{i \alpha j \beta}$,
\beq
f^{(x)}_{i \alpha j \beta}= - \frac{\partial t_{i \alpha j \beta}}{\partial x_{i} } 4 \sqrt{q_i}\langle  c_{i\alpha \sigma}^\dagger c_{j \beta \sigma} \rangle
\sqrt{q_j} \;.
 \eeq
The factor 4 arises from the two-fold spin degeneracy  and the Hermiticity. For the interacting case ($U \neq 0$), $\sqrt{q_i}$ or $\sqrt{q_j}$ are less than one
for $f$-orbitals but equal to one for $d$-orbitals. For the non-interacting case ($U=0$), the above formula is obtained by setting all
$q=1$. It can be shown that forces from the hybridization origin are always attractive.

The computed forces Eq.~(\ref{eq:fxx}) are then inserted into New's equation of motion (EOM) for each atom. The positions are advanced in time by a time step $\delta t$ by numerically integrating the EOMs with the Verlet algorithm. The resulting new atomic positions are then taken as input into Eq.~(\ref{Heff-model}), and new atomic forces Eq.~(\ref{eq:fxx})  are computed. The MD trajectory consists of the string of many time steps iterating back and forth through this two-step process.

\begin{figure}[t]
\includegraphics[scale=0.4]{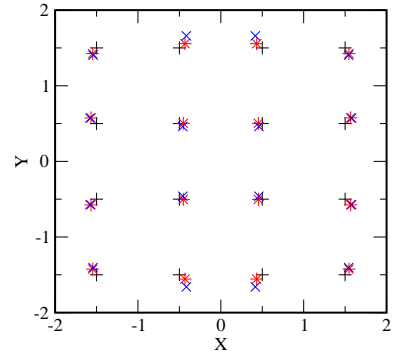}
\caption{(Color online) Comparison of atomic positions  for the second moment method (red `$\ast$') and exact diagonalization (blue `$\times$') for the non-interacting case. Initial positions  (black `$+$') are also plot.}
\label{Fig2}
\end{figure}

\subsection{Results}
As a  demonstration, consider a two-dimensional system that contains 16 atoms.  In the calculation, we take $t_{dd,0}=-1t$, $t_{df,0}=0.5t$, and $\Lambda_{0}=0.4t$. Hereafter all energies are measured in units of $t$. The bare $f$ level is chosen to be $\epsilon_{if\sigma}^{0}=-U/2$. The initial condition is 16 atoms forming a regularly spaced 4 x 4 array with a unit nearest-neighbor distance.
To benchmark our method and see the efficiency of second moment plus accurate approximate solution for double occupancy, we also performed the calculation based on exact diagonalization.
Since we are interested in finding only the equilibrium structure of the system, the velocity on each atom is set to zero before advancing by $\Delta t$ the numerical solution of the EOM. The resulting MD ``trajectory'' eventually finds a local minimum on the energy landscape as the atomic positions are converged and the residual forces are driven to the noise limit.
We started with the non-interacting case ($U=0$). Figure~\ref{Fig2} shows initial and  final positions of   of atoms after 5 millions of MD time steps i.e., sufficient to converge and where residual forces can be considered as noise.
The equilibrium structures obtained from the second moment and the exact diagonalization methods are quite similar with the root mean square deviation $\Delta r_{\text{RMS}}=\sqrt{\sum_{i=1}^{N}(\mathbf{r}^{\text{sm}}_{i}-\mathbf{r}^{\text{ED}}_{i})^2/N}$ of only about 0.058 unit distance. The CPU time in the case of second moment method is a factor of 20 times faster than the exact diagonalization. The second moment approach scales linearly with the number of atoms $N$ whereas ED scales as $N^3$. Therefore, if we had studied 10 times more particles, i.e., 160,  the increase speed factor would have been around 2000. The combination of second moment for approximate electronic structure quantities and a fast  Gutzwiller solver really opens the realm of possibilities for the simulations of large systems,  where electronic correlations play an important role.

\begin{figure}[t]
\includegraphics[scale=0.4]{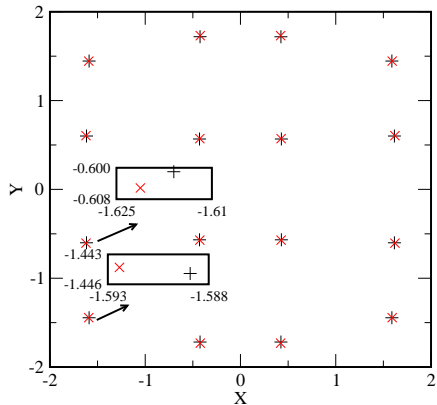}
\caption{(Color online) Comparison of atomic positions after MD process for second moment method for $U=0$ non-interacting case (black `$+$') and for $U=4t$ Gutzwiller interacting case (red `$\times$'). The expansion is small but can be seen in the zoom-in insets.}
\label{Fig3}
\end{figure}

We then started from the converged atomic positions as shown in Fig.~\ref{Fig2} and repeated the simulation with  a local Coulomb repulsion $U=4t$. The comparison between $U=0$ and $U=4t$ for the second moment approach is displayed in Fig.~\ref{Fig3}.
As expected and in accordance to our previous experience with Pu-$\delta$,~\cite{Julien2006} we see that the Gutzwiller $q$-factors have the effect of reducing the hybridization and the resultant attractive forces, which leads to  a slightly more expanded equilibrium structure as shown in the zoom-insets of Fig.~\ref{Fig3}. The same trend was also observed in the results (not shown here) obtained by exact diagonalization with $U=0$ versus $U=4t$.  We note that the structure expansion is quite small. The reason lies in the fact that in the Anderson-like model, the attractive force is contributed not only from the $d$-$f$ hybridization hopping but also significantly from the direct $d$-$d$ hopping. In addition, the efficiency of the hybridization reduction in the present model is proportional to $\sqrt{q_i}$. It is in contrast to the one-orbital Hubbard model, where the effective hopping integrals are proportional to $\sqrt{q_i} \sqrt{q}_j$.

\section{Summary and conclusion}
\label{sec:summary}
We have  derived for the first time a real-space version of the Gutzwiller method embedded into the MD simulations for strongly correlated electron systems. From the positions of a configuration of atoms, a Hamiltonian can be constructed in terms of hopping integrals, on-site energies and Coulomb repulsion terms. It is precisely these interaction terms that require treatment beyond mean-field HF theories, which allows for Gutzwiller method. This method is a variational method in the  Rayleigh-Ritz sense, for which one minimizes the energy of the system via a set of local Gutzwiller variational parameter dependent on each site, thereby providing an approximate ground state energy for the system. This minimization can be computationally demanding, especially when all the sites are inequivalent. This has motivated the development of an accurate analytical solution for finding the optimized double occupancy on each site.

MD simulation is a repeated two step process.  First, employing a Hellmann-Feynman theorem within the Gutzwiller ground state, we can calculate the quantum origin of the forces acting on the atoms.  This ground state energy defines an interatomic potential, which explicitly accounts for correlation effects. Second, the atomic forces derived from this interatomic potential are input into the classical equations of motion and the atomic positions are evolving forward in time for one time step using a numerical integrator (e.g., the Verlet algorithm). The step one is then repeated, where the new atomic positions define a new Hamiltonian whose new ground state can be found by the Gutzwiller method and so on.

A further approximation consists of avoiding exact diagonalization of the Hamiltonian from which, in principle, we can calculate the local densities of states (DOS) to obtain all necessary integrated quantities.  Instead, we have proposed to use the second moment approximation, in which the true DOS is replaced by a rectangular approximation having the same second moment.  Because the needed quantities to construct the variational ground state energy are basically integrated from DOS, they are less sensitive to the detailed structure of the DOS, thus validating the second moment approximation. The second moment of the energy is easily computed from a few Hamiltonian matrix elements, where we can set up the MD process without invoking exact diagonalization to solve iteratively the Gutzwiller minimization.  Therefore, a very accurate approximate, but analytical solution, is available, making this process feasible.
We concluded this first study with an application to a realistic case to show the potential of this approach, which to our knowledge, has never been developed and will open up  new possibilities for simulations of correlated electron materials with molecular dynamics. We have applied the Gutzwiller Molecular Dynamics method to the one correlated orbital per site case as exemplified in
 the lattice Anderson model. The generalization of these ideas to multiple correlated orbitals case is not
 made directly because of higher degeneracy. That is, the number of variational parameters increases as the number of atomic configurations (namely as $2^G$, with $G$ the degeneracy of the level)  and therefore  the number of local equations to be solved increases accordingly.
We defer to future work on how to reduce the number of variational parameters, and as in the one correlated orbital per site model, to find an analytical approach to the problem.

\begin{acknowledgments}
We thank S. Valone, W. A. Harrison, and J. M. Wills for useful  discussions. J.-P.J. would like to thank the Los Alamos National Laboratory  for the hospitality and financial support during his visits.
This work was carried out under the auspices of
the National Nuclear Security Administration of the U.S.
Department of Energy at LANL under
Contract No. DE-AC52-06NA25396, and was supported by the LANL ASC Program (J.D.K. \& J.-X.Z.). This work was in part supported by the Center for Integrated Nanotechnologies, a U.S. DOE user facility.
Some  preliminary results have been reported on the 2014 CECAM Workshop on Gutzwiller Wave Functions and Related Methods.
\end{acknowledgments}

\appendix*
\section{Second moment approximation for bond quantities}
We follow closely the derivation by Pettifor and co-workers~\cite{APSutton:1988,DGPettifor:1989} for the bond order in a tight-binding model.
To compute forces from the derivation of Hamiltonian, one needs average values like
$\langle c^{\dag}_ic_j\rangle$, with $i$ and $j$ being a short hand for site and spin-orbital states. Within
these notations, and for the purpose of demonstration, we write the Hamiltonian $H$ in simple tight-binding form:
\beq
H=\sum_{i \neq j} t_{ij}c_{i}^\dagger c_{j} + \sum_{i} \epsilon_{i}n_{i\sigma}\;,
\eeq
where on-site energies $\epsilon_{i}=\langle i|H|i\rangle$ can be identified as average value of the Hamiltonian on local state labelled by $i$  whereas hopping integrals
$t_{ij} =\langle i|H|j\rangle$ couple  states $i$ to $j$.
 The bracket is the thermal average obtained for
a general operator $O$ by
\beqn
\langle O\rangle= Tr \frac{e^{-\beta(H-\mu
N)}O}{Z}\;,
\eeqn
where $H$, $\mu$, $N$ and $Z$ are respectively the
Hamiltonian, the chemical potential, the operator number of
particles and the grand partition function. This average reduces
to the  ground state mean value at zero temperature.  From exact diagonalization (as we do for the cluster example developed here), these quantities can easily be calculated from the weights $w(j,n)$ of state $j$ (and similarly $i$) on  eigenstate labelled by $n$ and of energy $\epsilon_{n}$:
\beq
 \langle c^{\dag}_ic_j\rangle=\sum_{n} f(\epsilon_{n}) w^{*}(i,n)w(j,n)\;,
 \eeq
where $f(\epsilon_{n})$ is the Fermi distribution.

For large systems, where the speed of calculation is a limiting factor, it might be desirable however to avoid this diagonalization, and to find an approximate but cheap way to get them.
In a alternative route, they are obtained from the retarded Green function
(with Zubarev notation~\cite{Zubarev60})  $\mathcal{G}_{i,j}(\omega)\equiv \langle\langle c_j;c^{\dag}_i\rangle\rangle_{\omega}$, which is the Fourier
transform of $\mathcal{G}_{i,j}(t)=-i\theta(t)\langle \{c_j(t),c^{\dag}_i(0)\} \rangle$ for the
operators $c_j$ and $c^{\dag}_i$.
$\mathcal{G}_{i,j}(\omega)$ can be considered as an off-diagonal element of the
Green function, and in the case of effective independent electrons (with only 1-body operators, as it is the case for effective Hamiltonians) it reduces to the usual resolvent:
\beq
(\omega-H) \mathcal{G}=I \;,
\eeq
The diagonal element (\textit{i.e.} same states $i$ and $j$) relates to the $i$-projected density of states $N_i(\omega)$ via:
\beq N_i(\omega)=- \frac{1}{\pi} \text{Im} \mathcal{G}_{i,i}(\omega) \label{dos}\;.
\eeq
There are several useful relations and sum rules that fulfills the Green
function (see Ref.~\onlinecite{Zubarev60} for demonstration):

\beqn
-\frac{1}{\pi} \int \text{Im} \langle \langle c_j;c^{\dag}_i \rangle \rangle _{\omega}d\omega&=&\langle \{c_j,c^{\dag}_i\}\rangle \nonumber\\
&=&\delta_{i,j}\;,
\label{sumrule1}
\eeqn 
and the so-called spectral theorem provides a direct way
to compute the average values we are looking for:
\beqn -\frac{1}{\pi} \int f(\omega) \text{Im}
\langle \langle c_j;c^{\dag}_i\rangle\rangle_{\omega}d\omega&=&\langle c^{\dag}_i c_j\rangle \;.
\label{spectraltheorem}
\eeqn

When there is no magnetic field, the following relation holds:
\beqn
\langle \langle c_i;c_j^{\dag}\rangle\rangle=\langle \langle c_j;c^{\dag}_i\rangle\rangle\;.
 \label{nomagfieldgreen}
 \eeqn

The main idea in the present approximate calculation is the following: we express exactly the off-diagonal element of Green function  as a linear combination of diagonal elements of Green functions of bonding and anti-bonding states,created by $c^{\dag}_{B\,or\, A}=\frac{c^{\dag}_i\pm c^{\dag}_j}{\sqrt{2}}$.
Indeed, one can easily show
\beq
 \mathcal{G}_{i,j}=\frac{1}{2}(\mathcal{G}_{B,B}-\mathcal{G}_{A,A})\;,
\eeq
from which the desired quantity is obtained thanks to relations (\ref{spectraltheorem}), (\ref{dos}) and  (\ref{nomagfieldgreen}):
\beq
\langle c^{\dag}_i c_j\rangle =-\frac{1}{2\pi} \int f(\omega)( N_{B}(\omega)-N_{A}(\omega))d\omega\;.
\label{cdagc2m}
\eeq

Then the second moment approximation is applied to the DOS --the imaginary part of Green function-- calculated for bonding and anti-bonding states, respectively. The  second moment approximation is based on the constraints that all necessary quantities are integrated quantities. Consequently, they are not sensitive to the fine details of the DOS, which will be replaced by rectangular DOS having the same second moment that the true ones. The rectangular $i$-projected (centered on site energy $\varepsilon_i$ with width $W_i$ and height $1/W_i$) DOS has its second moment given by: $\varepsilon^2_i + W^2_i/12$ whereas a direct path-counting (see Ref.~\onlinecite{Cyrot-Lackmann73})  gives $\varepsilon^2_i + \sum_{j\neq i} t^2_{ij}$. Identification between those two relations fixes uniquely the bandwidth $W_i$, from which the approximated $i$-projected DOS can be computed. This procedure has been widely used in semi-empirical molecular dynamics as in
Ref.~\onlinecite{Ackland2003}, for approximate local DOS. The extension we suggest here is to used it also for the bonding and anti-bonding DOS, $ N_{B}(\omega)$ and $N_{A}(\omega)$, to obtain the bond quantities.

It should be noted that in the particular case of the second moment of the projected density of states on A (or B), the second moment defined by the mean value $\langle A|H^{2}|A\rangle$ provides unusual terms as follows:
\beq
\mu_{2,ij}=\langle i|H^{2}|j\rangle=\sum_{k}\langle i|H|k\rangle\langle k|H|j\rangle\;.
\eeq
Such terms do not appear in the traditional way of calculating the moments starting from a given orbital at a specific site $i$, where the second moment is simply given by the number of paths  starting from $i$, returning after two jumps back to the same site: this later simpler case is obtained by setting $i=j$ in the last equation.
Finally, we get the second moment on either $\mu_{2,AA}$ or $\mu_{2,BB}$ combination:
\begin{subequations}
\begin{equation}
\mu_{2,AA} = \langle A|H^{2}|A\rangle=\frac{1}{2} (\mu_{2,ii}+ \mu_{2,jj}- 2 \Re \mu_{2,ij})  \;,
\end{equation}
\begin{equation}
\mu_{2,BB} = \langle B|H^{2}|B\rangle=\frac{1}{2} (\mu_{2,ii}+ \mu_{2,jj}+ 2 \Re \mu_{2,ij})\;,
\end{equation}
\end{subequations}
together with the center of band $\epsilon_A$ (resp. $\epsilon_B$ ):
\begin{subequations}
\begin{eqnarray}
\epsilon_A &=& \langle A|H|A\rangle=\epsilon_i+\epsilon_j - 2 \Re t_{ij}  \;,\\
\epsilon_B &=& \langle B|H|B\rangle=\epsilon_i+\epsilon_j + 2 \Re
t_{ij} \;,
\end{eqnarray}
\end{subequations}
from which we can obtain the related bandwidth $W_A=\sqrt{12(\mu_{2,AA}-\epsilon^2_A)}$ (similar expression holds for $W_B$).
Finally using expression (\ref{cdagc2m})  for rectangular DOS $N_A$ and $N_B$, we have the desired quantity:
\beq
 \langle c^{\dag}_ic_j\rangle=\frac{E_F-\epsilon_B}{2 W_B}-\frac{E_F-\epsilon_A}{2 W_A}\;,
\eeq
where $E_F$  is the chemical potential at zero temperature. All this demonstration can be straightforwardly be extended to multiband case adding orbital index and spin
to labels $i$ and $j$. This procedure presents the great advantage of being very rapid compared to exact diagonalization and one can check that the Green function related to approximate DOS also fulfills sum rule as given in Eq.~(\ref{sumrule1}).

\end{document}